\documentclass[useAMS,usenatbib]{mn2e}
\usepackage{graphicx,amssymb, epstopdf}
\newcommand{\vc}[1]{\bmath{#1}}
\newcommand{\uvc}[1]{\hat{\bmath{#1}}}

\newcommand{\beq}{\begin{equation}}
\newcommand{\eeq}{\end{equation}}
\newcommand{\bea}{\begin{eqnarray}}
\newcommand{\eea}{\end{eqnarray}}
\newcommand{\lt}{\left}
\newcommand{\rt}{\right}

\newcommand{\Secref}[1]{Section~\ref{#1}}

\newcommand{\eqref}[1]{equation~(\ref{#1})}

\newcommand{\dd}{\partial}

\newcommand{\bnabla}{\vc {\nabla}}

\newcommand{\vths}{v_{\rmn{th},s}}
\newcommand{\vthi}{v_{\rmn{th},i}}

\newcommand{\g}{\gamma}
\newcommand{\e}{\epsilon}
\newcommand{\E}{\varepsilon}
\newcommand{\pd}{p'}
\newcommand{\gd}{\gamma'}
\newcommand{\ggd}{\g\gd}
\newcommand{\D}{\delta}
\newcommand{\para}{\parallel}
\newcommand{\mec}{m_e c^2}

\newcommand{\dsigmaperp}{d^2\sigma_{\perp}}
\newcommand{\dsigmapara}{d^2\sigma_{\para}}

\newcommand{\dsigmasum}{d^2\sigma_{0}}
\newcommand{\dsigmadiff}{d^2\sigma_{1}}

\newcommand{\ppars}{p_{\parallel s}}
\newcommand{\pperps}{p_{\perp s}}

\title[Polarization of bremsstrahlung emission]{Polarization of thermal bremsstrahlung emission due to electron pressure anisotropy}
\author[S. V. Komarov, I. I. Khabibullin, E. M. Churazov  and A. A. Schekochihin]{
S. V. Komarov$^{1,2}$\thanks{E-mail: komarov@mpa-garching.mpg.de}, I. I. Khabibullin$^{1,2}$, E. M. Churazov$^{1,2}$  and A. A. Schekochihin$^{3,4}$\\
$^{1}$Max Planck Institute for Astrophysics, Karl-Schwarzschild-Strasse 1, 
85741 Garching, Germany\\
$^{2}$Space Research Institute (IKI), Profsouznaya 84/32, Moscow 117997, Russia\\
$^{3}$The Rudolf Peierls Centre for Theoretical Physics, University of Oxford, 
1 Keble Road, Oxford OX1 3NP, United Kingdom\\
$^{4}$Merton College, Oxford OX1 4JD, United Kingdom
} 

\begin{document}
\maketitle
\label{firstpage}

\begin{abstract}
Astrophysical plasmas are typically magnetized, with the Larmor radii 
of the charged particles many orders of magnitude smaller than their 
collisional mean free paths. The fundamental properties of such plasmas, 
e.g., conduction and viscosity, may depend on the instabilities driven 
by the anisotropy of the particle distribution functions and operating at 
scales comparable to the Larmor scales. We discuss a possibility that 
the pressure anisotropy of thermal electrons could produce polarization 
of thermal bremsstrahlung emission. In particular, we consider  
coherent large-scale motions in galaxy clusters to estimate the level 
of anisotropy driven by stretching of the magnetic-field lines by 
plasma flow and by heat fluxes associated with thermal gradients. Our 
estimate of the degree of polarization is $\sim 0.1 \%$ at energies $\gtrsim 
kT$. While this value is too low for the forthcoming generation of 
X-ray polarimeters, it is potentially an important proxy for the 
processes taking place at extremely small scales, which are impossible to 
resolve spatially. The absence of the effect at the predicted level may 
set a lower limit on the electron collisionality in the ICM. At the same time, 
the small value of the effect implies that it does not preclude the use of clusters as (unpolarized) 
calibration sources for X-ray polarimeters at this level of accuracy.

\end{abstract}


\section{Introduction}
\label{sec:intro}

For the electrons in a hot tenuous astrophysical plasma, the equilibration time 
scale due to Coulomb collisions is often sufficiently long compared to other 
characteristic time scales to allow for deviations from thermal equilibrium 
manifested by anisotropies or non-thermal tails. The latter typically require a 
mechanism to accelerate a fraction of particles to high energies, e.g., 
magnetic reconnection or diffusive shock acceleration (\citealt{Krymsky1977,Axford1977,
Blandford1978,Bell1978}). Anisotropies, on the other hand, are commonly associated 
with the presence of a magnetic field. 

In many astrophysical plasmas, magnetic fields are strong enough to force a 
charged particle to orbit around a field line with the Larmor radius many orders 
of magnitude smaller than the particle's collisional mean free path. If the magnetic 
field is not constant in time, adiabatic invariance compels the perpendicular 
and parallel components of the particle's velocity to adjust to the field magnitude 
in different ways, thus 
producing pressure anisotropy \citep{CGL,Kulsrud1964}. A heat flux along the field lines also contributes 
to anisotropy. Above a certain threshold, pressure anisotropies trigger kinetic 
microinstabilities, e.g., firehose and mirror \citep{Chandrasekar1958,Parker1958,Hasegawa}, which are believed to 
hold the anisotropy at a marginal level by increasing the effective collision 
rate via scattering off magnetic perturbations and magnetic trapping (for 
observational evidence in the solar wind, see \citealt{Kasper2002,Hellinger2006,
Bale2009}; for theoretical discussion, see, e.g., \citealt{Kunz2014,Melville2015} and 
references therein). 

Temporal changes of magnetic-field strength may be caused by random turbulent 
motions or by a specific ordered plasma flow, e.g., a flow past a cold dense cloud of 
gas in ``cold fronts'' in the ICM \citep[see, e.g.,][for a review]{MarkevitchVikhl2007} 
or a shear flow in accretion disks \citep{Sharma2006}. Cold fronts also manifest sharp temperature 
gradients at the interface between the cold cloud and the hot ambient plasma. 
In a hot rarefied plasma, the electron temperature anisotropy generated by 
both the magnetic-field evolution and heat fluxes leaves an imprint in the 
form of polarization of bremsstrahlung emission \citep[for an example in solar 
flares, see, e.g.,][]{Haug}. If a flow orients the magnetic field in some 
preferential direction, the polarization does not cancel out and potentially can 
be observed by X-ray polarimeters.

In this paper, we examine the possible magnitude and detectability of 
electron pressure anisotropy in galaxy clusters. 
We start by describing the theoretical framework for the problem at hand. 
Generation of pressure anisotropies in a plasma with evolving magnetic fields 
and temperature gradients is discussed in \Secref{sec:anis}. In \Secref{sec:pol}, 
we derive the polarization of bremsstrahlung emission for a given anisotropic 
bi-Maxwellian electron distribution. Then we proceed with application of our 
theory to cold fronts and shocks in the ICM with the help of analytical models 
and numerical MHD simulations (Section~\ref{sec:cf}). The effects are weak, 
of order $0.1\%$, but they produce a characteristic pattern and may provide 
constraints on the pressure anisotropy and electron collisionality in the ICM. We briefly discuss the role of 
the effects for observations of galaxy clusters 
with future X-ray calorimeters in Section~\ref{sec:disc}. 
Finally, we summarize our findings in Section~\ref{sec:concl}.

\section{Theoretical framework}
\subsection{Generation of pressure anisotropies in a weakly collisional plasma}
\label{sec:anis}

In an astrophysical plasma, the Larmor radii $\rho_s$ of all particle species are 
typically much smaller 
than their collisional mean free paths $\lambda_s$; equivalently, their 
collision frequencies $\nu_s$ are much smaller than the Larmor frequencies 
$\Omega_s$ ($s=i,e$, with $e$ the electrons, $i$ the ions). If the magnetic-field 
strength $B$ changes slowly, $\gamma = B^{-1}dB/dt$, each particle conserves 
its first adiabatic invariant $\mu = v_{\perp s}^2/2B 
\propto p_{\perp s}/B$, which is the magnetic moment of the particle, where 
$v_{\perp s}$ is the component of the particle's velocity perpendicular to 
the field line, $p_{\perp s}$ is the perpendicular pressure. To demonstrate
how pressure anisotropy is driven and sustained by evolving magnetic fields 
in an incompressible plasma with no heat flux, we express the assumption that 
$\mu$ is conserved but for rare occasional collision as
\beq
\frac{1}{p_{\perp s}} \frac{d p_{\perp s}}{d t} \sim \frac{1}{B} \frac{d B}{d t}- 
\nu_s \frac{p_{\perp s}-p_{\parallel s}}{p_{\perp s}},
\eeq      
where $p_{\parallel s}$ is the parallel pressure, and the last term corresponds 
to isotropization of pressure by collisions. If $\gamma\ll\nu_s$, the pressure anisotropy 
$\Delta_s$ can be estimated from the balance between the collisional relaxation 
and the rate of change of the magnetic field:
\beq
\Delta_s \equiv \frac{p_{\perp s}-p_{\parallel s}}{p_s} \sim \frac{1}{\nu_s} 
	\frac{1}{B} \frac{d B}{dt}=\frac{\gamma}{\nu_s}.
\eeq
It is clear from this estimate that the electron 
anisotropy is $\nu_e/\nu_i \sim (m_i/m_e)^{1/2} \approx 40$ times weaker 
than that of the ions (if the electron and ion temperatures are equal).

The more general form of $\Delta_s$ taking into account the evolution 
of the parallel pressure can be obtained from the so-called CGL equations 
\citep{CGL} with collisions retained, which are derived by taking 
second moments of the kinetic magnetohydrodynamics equations 
(KMHD). The KMHD equations arise after averaging the full kinetic equation 
over the gyroangle. The CGL equations read \citep{Schek2010}
\bea
\nonumber
\pperps \frac{d}{d t} \ln{\frac{\pperps}{n_s B}} &=& 
		\bnabla \cdot (q_{\perp s} \vc{b}) 
		- q_{\perp s} \bnabla \cdot \vc{b}\\		 
		&&\,- \nu_s (\pperps - \ppars),
\label{eq:cgl1}\\
\nonumber
\ppars \frac{d}{d t} \ln{\frac{\ppars B^2}{n_s^3}} &=& 
		\bnabla \cdot (q_{\parallel s} \vc{b})
		+ 2 q_{\perp s} \bnabla \cdot \vc{b}\\ 
		&&\,- 2 \nu_s (\ppars - \pperps),
\label{eq:cgl2}
\eea
where $d/dt=\dd/\dd t + \vc{u}_s \cdot \bnabla$ is the convective derivative 
associated with species $s$ ($\vc{u}_s$ is the plasma flow velocity), 
$\vc{b}$ is the unit vector in the direction of the magnetic 
field, $n_s$ the number densities, 
\bea
\pperps &=& \int d^3 \vc{w} \frac{m_s w_{\perp}^2}{2} f_s,\\
\ppars &=& \int d^3 \vc{w} m_s w_{\parallel}^2 f_s
\eea
are the perpendicular and parallel pressures,
\bea
q_{\perp s} &=& \int d^3 \vc{w} \frac{m_s w_{\perp}^2}{2} w_{\parallel} f_s,\\
q_{\parallel s} &=& \int d^3 \vc{w} \ m_s w_{\parallel}^3 f_s,
\eea
$q_{\perp s}$ and $q_{\parallel s}$ are heat fluxes (the parallel 
flux of the ``perpendicular internal energy'' and the parallel flux of 
the ``parallel internal energy'', respectively), $\vc{w}$ the thermal component 
of a particle's velocity, $f_s$ the distribution functions of the particles. 
Subtracting equation~(\ref{eq:cgl2}) from equation~(\ref{eq:cgl1}), we 
get an evolution equation for the pressure anisotropy:
\bea
\nonumber
\frac{d}{d t} (\pperps - \ppars) &=& 
	(\pperps + 2\ppars) \frac{1}{B}\frac{d B}{d t}		
	+ (\pperps - 3\ppars) \frac{1}{n_s}\frac{d n_s}{d t} \\
\nonumber
	&&- \bnabla \cdot [(q_{\perp s} - q_{\parallel s}) \vc{b}] 
	- 3 q_{\perp s} \bnabla \cdot \vc{b}\\ 
	&&- 3 \nu_s (\pperps - \ppars).	 
\eea
Assuming that collisions are fast compared to the fluid motions, 
the pressure anisotropy is then small, $\ppars - \pperps \ll \pperps 
\approx \ppars \approx p_s$, the collisional heat fluxes are 
$q_{\perp s} \approx (1/3) q_{\parallel s}$, 
and the total heat flux along a field line 
$q_s=q_{\perp s}+q_{\parallel s}/2=(5/6) q_{\parallel s}$. 
The value of the anisotropy is set by the 
balance between collisional relaxation and various driving terms:
\bea
\nonumber
\Delta_s \equiv \frac{\pperps - \ppars}{p_s} &\approx& 
			\frac{1}{\nu_s} \bigg [ \frac{1}{B} \frac{d B}{d t}  
		   -\frac{2}{3}\frac{1}{n_s}\frac{d n_s}{d t} \\
 &&+ \frac{4 \bnabla\cdot (q_s \vc{b}) - 6 q_s \bnabla \cdot \vc{b}}
 						{15 p_s} \bigg ] .
\label{eq:tot_anis}
\eea
Thus, the pressure anisotropy is driven by changing magnetic-field 
strength, changing particle density, and by parallel heat fluxes. 

It is useful to estimate the degree of anisotropy induced by different 
driving terms in \eqref{eq:tot_anis}. If we consider 
fluid motions with velocity $u$ at scale $L_u$, variations of $B$ at the 
scale of the velocity field $L_B=L_u$, and parallel temperature 
gradient $\nabla_{\parallel} T_s \sim \delta T_s/L_T$ at scale $L_T$, we 
can evaluate the contribution $\Delta_{B,n;s}$ of changing $B$ and $n$, 
and the contribution $\Delta_{T;s}$ of the heat fluxes, to the total anisotropy as
\bea
\label{eq:DB}
\lefteqn{\Delta_{B,n;s} \sim \frac{u}{\vths} \frac{\lambda_s}{L_u},}\\
\label{eq:DT}
\lefteqn{\Delta_{T;s} \sim \frac{\lambda_s^2}{L_T L_u} \frac{\delta T_s}{T_s},}
\eea
where we have used the expression for the heat flux $q_s = - \kappa_s 
\nabla_{\parallel} T_s$ with thermal conductivity $\kappa_s \sim n_s 
\vths \lambda_s$, $\lambda_s$ is the mean free path, $\vths$ the 
thermal speed. Assume that the flow velocity is nearly sonic, 
$u \sim \vthi$, and that the variations of temperature are 
of order unity, $\delta T / T \sim 1$. Then $\Delta_{T} 
\sim \lambda^2 / (L_u L_T)$ for both particle species, and 
$\Delta_{B,n;s} \sim \lambda / L_u \times \vths/\vthi$. 
Hence, in our ordering, for the ions, the term linked to 
the magnetic-field changes $\Delta_{B,n;i} \sim \lambda / 
L_u$ is dominant if $L_T \gg \lambda$ (even in astrophysical 
systems with very sharp temperature gradients, e.g., cold 
fronts or buoyant bubbles of relativistic plasma in the ICM, 
the magnetic-field lines are typically stretched by the 
fluid flow in the direction perpendicular to the gradient 
\citep[e.g.,][]{Komarov2014}, thus significantly increasing 
the scale of temperature variation along the field lines). 
For the electrons, $\Delta_{B,n;e} \sim 1/40 \times \lambda 
/ L_u$, and the two contributions can be of the same order 
($\Delta_{B,n;e} \sim \Delta_{T;e}$), 
depending on the properties of the flow and the orientation 
of the magnetic-field lines connecting the hot and cold regions 
of the plasma. Note that the total anisotropy is bounded from 
below by the firehose instability, $\Delta_e+\Delta_i > -2/\beta$, 
where $\beta$ is the ratio of thermal to magnetic-energy densities. 
If the ion and electron anisotropies are small ($\Delta_i,~\Delta_e \ll 1$), 
the mirror instability also compels the total anisotropy to stay 
below the mirror marginal level, $\Delta_e+\Delta_i \lesssim 1/\beta$.  
Therefore, in regions of high plasma 
$\beta$, either $\gamma$ ($=B^{-1}dB/dt$) or $\nu_s$ is modified by the instabilities to 
keep the anisotropy between the marginal levels, $-2/\beta < 
\Delta_e+\Delta_i < 1/\beta$ \citep[e.g.,][]{Kunz2014,Melville2015}. 
In Appendix~\ref{app:A}, we calculate the total anisotropy for 
the simulated cold fronts (Section~\ref{sec:cf}) and mark 
the regions where firehose and mirror instabilities 
could develop. Because in our work the ion anisotropy is typically 
dominant, the two instabilities are regulated by the ions. 

We are primarily interested in electron pressure anisotropy 
because of its possible observational imprint in the form of polarization of 
thermal bremsstrahlung. From the above estimates, it is clear that in the 
case of astrophysical systems with large temperature gradients, the driving 
term linked to heat fluxes must be taken into account along with the 
driving by the magnetic-field changes. We do this in detail for cold 
fronts in \Secref{sec:cf}.    
 
\subsection{Polarization of thermal bremsstrahlung by electron anisotropy}
\label{sec:pol}

Consider first the polarization of bremsstrahlung emission from an electron 
beam deflected by a single ion. At low energies (compared to the kinetic 
energy of an electron), photons produced by small-angle scattering of the 
electrons off the ion are polarized in the plane perpendicular to the electron 
beam due to the mainly perpendicular acceleration that slightly changes 
the direction of the electron velocity. At higher energies, when both the 
direction and magnitude of the electron velocity change significantly, 
polarization becomes dominated by the acceleration the electrons experience 
parallel to the beam. Below, we demonstrate that the latter regime is of 
first importance for our problem, because the degree of polarization is 
considerably larger at high energies in the case of thermal bremsstrahlung 
from a cloud of anisotropic electrons.

\begin{figure}
\centering
\includegraphics[width=84mm]{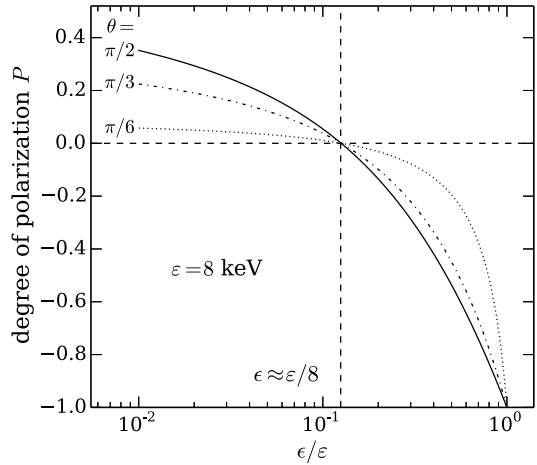}
\caption[The degree of bremsstrahlung polarization from a beam of 
electrons]
{The degree of bremsstrahlung polarization $ P(\e,\theta)=
(\dsigmaperp - \dsigmapara)/(\dsigmaperp + \dsigmapara)$ from a beam of 
electrons of energy $\E=8$ KeV as a function of the emitted photon's energy 
$\e$ and the angle $\theta$ between the beam axis and the photon's 
momentum.}
\label{fig:beam}
\end{figure}

Bremsstrahlung emission from a beam of electrons of kinetic energy $ \E $  is 
fully described by differential cross sections per unit solid angle 
and photon energy, $\dsigmaperp(\E,\e,\theta)$ and $\dsigmapara(\E,\e,
\theta)$, for the components perpendicular and parallel to the radiation 
plane (spanned by an emitted photon's and an initial electron's momenta). 
Here, $d^2 =d / ( d \e d \Omega)$,  $\e$ stands for the emitted photon's 
energy, and $ \theta $ for the angle between the emitted photon's momentum 
and the beam axis. We use the fully relativistic cross sections calculated 
by \cite{Gluckstern} in the first Born approximation, which are appropriate 
for the problem at hand \footnote{The limit of validity of this 
approximation is given by condition $ \E'/m_e c^2 \gg (Z/137)^2~$, 
where $Z$ is the charge of the scattering ion in atomic units, $\E'$ 
the energy of an outgoing electron \citep{Gluckstern}. For $Z=1$, the 
condition is satisfied for outgoing electrons at energies $\E' \gg 30$ 
eV.}. Because the formulae in the original paper by \cite{Gluckstern}, 
as well as those given later by \cite{BaiRamaty}, are both replete with 
typos, we provide the correct explicit expressions for the cross sections 
in Appendix~\ref{app:B}. The degree of polarization is $P(\e,\theta)=
(\dsigmaperp - \dsigmapara)/(\dsigmaperp + \dsigmapara) = \dsigmadiff 
/ \dsigmasum$, where $\dsigmadiff$ is the differential cross section 
of the polarized emission, $\dsigmasum$ of the total emission. Its 
dependence on the photon energy and direction with respect to the beam 
axis is illustrated in Fig.~\ref{fig:beam}. The transition between the 
perpendicular and parallel polarization occurs at photon energy $\e 
\approx \E/8$. As noted before, the perpendicular polarization at low 
energies is produced by small-angle scattering of the electrons, 
while the parallel is the result of collisions that significantly 
change the electron energy. 

\begin{figure}
\centering
\includegraphics[width=84mm]{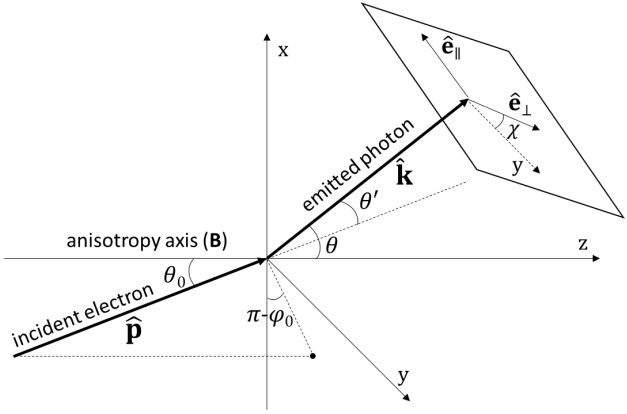}
\caption[Geometry for the problem of the polarization of 
bremsstrahlung emission from a cloud of electrons]{Geometry for the problem of the polarization of 
bremsstrahlung emission from a cloud of electrons.}
\label{fig:vectors}
\end{figure}

Since the differential cross sections presented above are essentially 
the ``Green's functions'' of the bremsstrahlung emission, the total and 
polarized emission from a cloud of electrons can be found by integrating 
over the electron distribution function. Let us introduce a spherical 
coordinate system and assume that the electron distribution is 
axisymmetric with respect to the magnetic-field direction, taken to be the $z$ axis. We denote the unit vector 
in the direction of the incoming electron $\uvc{p}=(\sin{\theta_0} 
\cos{\phi_0}, \sin{\theta_0} \sin{\phi_0}, \cos{\theta_0} )$, and the 
direction of the line of sight $\uvc{k} = (\sin{\theta}, 0, \cos{\theta})$ 
(choose $\phi=0$ without loss of generality because the resulting 
polarization pattern is also axisymmetric). The geometry of the vectors 
is illustrated in Fig.~\ref{fig:vectors}. The polarization directions 
perpendicular and parallel to the plane spanned by the vectors $\uvc{p}$ 
and $\uvc{k}$ (the radiation plane) are, respectively,
\bea
\uvc{e}_{\perp} &=& \frac{\uvc{p} \times \uvc{k}}{|\uvc{p} \times \uvc{k}|},\\
\uvc{e}_{\para} &=& \frac{\uvc{k} \times (\uvc{p} \times \uvc{k})}{|\uvc{p} 
\times \uvc{k}|}.
\eea
Then $\uvc{e}_{\perp}$ is rotated by angle $\chi$ (see Fig.~
\ref{fig:vectors}) with respect to the $y$ direction, which is the 
perpendicular polarization direction in the reference plane $xz$ that 
contains the line of sight $\uvc{k}$ and the symmetry axis $\vc{B}$. The angle $\chi$ is expressed as  
\beq
\label{eq:chi}
\cos \chi = \hat{e}_{\perp y} = (\sin \theta \cos \theta_0  - 
\cos \theta \sin \theta_0 \cos \phi_0 ) / \sin \theta',
\eeq 
where $\theta'$ is the angle between $\uvc{p}$ and $\uvc{k}$:  
\beq
\cos \theta' = \cos \theta \cos \theta_0  + \sin \theta \sin \theta_0 
\cos \theta_0.
\eeq
Linear polarization (for unpolarized electrons, bremsstrahlung 
photons are never circularly polarized) is described by the two 
independent Stokes parameters $P_1$ and $P_2$: $P_1$ corresponds to
the degree of polarization with respect to a given reference plane 
($xz$ in our case); $P_2$ to the degree of polarization with 
respect to a plane rotated around the line of sight by $\pi/4$ from 
the reference plane. For a given momentum of the initial electron, $P_1$ 
and $P_2$, normalized by the total intensity, are transformed by 
rotation of the radiation plane relative to the reference plane as
\bea
\nonumber
P_{1,\uvc{p}} &=& \cos{2\chi} \frac{\dsigmadiff}{\dsigmasum},\\
\label{eq:P1P2}
P_{2, \uvc{p}} &=& \sin{2\chi} \frac{\dsigmadiff}{\dsigmasum}.
\eea
Thus, knowing the expression for the angle $\chi$ [equation~(\ref{eq:chi})] 
between the radiation and reference planes, we can calculate the 
degree of polarization of bremsstrahlung emission from a cloud of 
electrons $P_{1,2} = \mathcal{I}_{1,2}/\mathcal{I}_0$, where 
$\mathcal{I}_{1,2}$ is the intensity of the polarized emission and 
$\mathcal{I}_0$ the total intensity, both integrated over the 
electron distribution $F(\E,\theta_0)$:
\bea
\nonumber
\mathcal{I}_0(\e,\theta)&=&n_i\int_{\e}^{\infty} d\E\int_{-1}^{1} d(\cos\theta_0) 
		\int_{0}^{2\pi} d\phi_0 \ \\
		&& \times \lt [ v(\E) F(\E,\theta_0)~ \dsigmasum (\E,\e,\theta') \rt ],\\
\label{eq:isum}
\nonumber 
\mathcal{I}_1(\e,\theta)&=&n_i\int_{\e}^{\infty} d\E\int_{-1}^{+1} d(\cos\theta_0) 
		\int_{0}^{2\pi} d\phi_0 \ \\
		&& \times \lt [ v(\E) F(\E,\theta_0) \cos 2\chi ~ \dsigmadiff (\E,\e,\theta') \rt ], \\
\label{eq:idiff}
\nonumber 
\mathcal{I}_2(\e,\theta)&=&n_i\int_{\e}^{\infty} d\E\int_{-1}^{+1} d(\cos\theta_0) 
		\int_{0}^{2\pi} d\phi_0 \ \\
		&& \times \lt [ v(\E) F(\E,\theta_0) \sin 2\chi ~ \dsigmadiff (\E,\e,\theta') \rt ].
\eea
Due to the axisymmetry of the electron distribution function, 
$\mathcal{I}_2$ integrates to zero (see, e.g., the appendix of 
\citealt{Haug} for a mathematical proof of this), and the total 
degree of linear polarization is $P = \mathcal{I}_1/\mathcal{I}_0$.
 
\begin{figure}
\centering
\includegraphics[width=84mm]{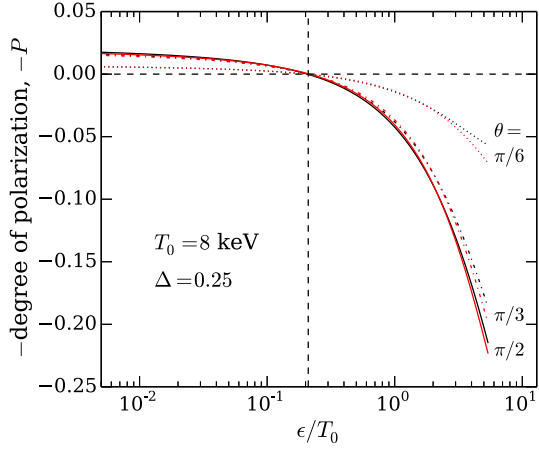}
\caption[The degree of bremsstrahlung polarization from a cloud of electrons with a bi-Maxwellian distribution]
{The degree of bremsstrahlung polarization from a cloud of 
electrons with a bi-Maxwellian distribution at temperature 
$T_0=8$~keV and anisotropy level $\Delta=0.25$ [\eqref{eq:Delta}] 
as a function of the emitted photon energy $\e$ and the angle between 
the axis of anisotropy and the line of sight. Results in the linear 
approximation [equations~(\ref{eq:fvalphap}) and (\ref{eq:f0})] are shown 
in red for comparison. The polarization degree is plotted with the 
minus sign to facilitate comparison with Fig.~\ref{fig:beam}. The 
opposite sign comes from the fact that the electron pressure 
anisotropy $\Delta$ is defined to be positive for $T_{\perp} > 
T_{\parallel}$ [equation~(\ref{eq:Delta})].}
\label{fig:pol_th} 
\end{figure} 

The distribution function $ F(\E,\theta_0)$ is related to the velocity 
distribution function $f(v,\theta_0)$ as
\smallskip
\beq
\label{eq:fefromfv}
F(\E,\theta_0)=v^2 \ f(v,\theta_0)\frac{dv}{d\E}.
\eeq
For the velocity distribution function, we employ a bi-Maxwellian: 
\bea
\nonumber
f(v,\theta_0)&=&n_e\left(\frac{m_e}{2\pi T_{\perp}}\right)\left(\frac{m_e}{2\pi T_{\para}}\right)^{1/2}\\ 
	&&\times \exp\left[-\frac{m_e v^2}{2 T_0}\left(\frac{T_0}{T_{\perp}}\sin^2\theta_0+\frac{T_0}{T_{\para}}\cos^2\theta_0\right)\right].
\label{eq:fv}
\eea
where $T_0=(1/3) T_{\para} + (2/3) T_{\perp}$ is the total temperature. If the anisotropy 
\beq
\label{eq:Delta}
\Delta \equiv \frac{T_{\perp}-T_{\para}}{T_0}
\eeq 
is small and $\Delta m_e v^2 / (2 T_0) \ll 1$, 
one can expand the distribution function to the first order in $\Delta$:
\beq
\label{eq:fvalphap}
f(v,\theta_0)=f_0(v)+\delta f_\Delta(v,\theta_0),
\eeq
where $f_0(v)$ is an isotropic Maxwell distribution at temperature $ T_{0}$:
\beq
\label{eq:f0}
f_0(v)=n_e\left(\frac{m_e}{2\pi T_{0}}\right)^{3/2}\exp\left(-\frac{m_e v^2}{2 T_0}\right)
\eeq
and the anisotropic perturbation is
\beq
\label{eq:df}
\delta f_\Delta(v,\theta_0)=\Delta ~\frac{m_e v^2}{2 T_0}~\left(\frac{1}{3}-\cos^2\theta_0\right) ~ f_0(v).
\eeq
Using equations~(\ref{eq:idiff}), (\ref{eq:fefromfv}), 
and (\ref{eq:df}), we obtain the degree of polarization of thermal 
bremsstrahlung for a small anisotropic perturbation of the electron 
distribution, when the linear approximation [expansion in $\Delta$ in equation~(\ref{eq:fvalphap})] is applicable:
\beq
\label{eq:linpoldeg}
P(\e, T_0,\theta) =  \Delta \sin^2 \theta \ G(\e, T_0),
\eeq
where $G(\e,T_0)$ becomes a function of $\e/T_0$ at temperatures $T_0 \lesssim 
10$~keV. At $\e \sim $ a few $T_0$, $G(\e,T_0) \sim 1$. The degree of 
polarization from a cloud of anisotropic electrons with $\Delta=0.25$ 
at $T_0=8$~KeV is shown in Fig.~\ref{fig:pol_th} in black for a 
general bi-Maxwellian distribution, and in red in the linear 
approximation [equation~(\ref{eq:linpoldeg})]. We see that the 
linear approximation holds at least up to $\Delta\approx 0.25$. 

\section{Application to cold fronts and shocks in the ICM}
\label{sec:cf}
\subsection{Qualitative estimates}
\label{sec:cf_qual}

Cold fronts are sharp discontinuities of temperature and density 
seen in X-rays in a number of clusters \citep{Markevitch2000, Ettori2000, 
Vikhlinin2001, MarkevitchVikhl2007}. These are commonly associated with 
a flow of the hot ambient ICM plasma around a cold subcluster moving in 
the host cluster nearly at the virial speed. The plasma flow produces 
draping of the frozen-in magnetic-field lines over the cold cloud 
\citep[e.g.,][]{Lyutikov2006,Asai2007,Dursi2008}. Near the front, the 
flow is essentially a convergence flow, and the field lines are 
continuously stretched along the interface. This leads to perpendicular 
orientation of the field lines and temperature gradient and likely 
inhibits thermal conduction, preserving the sharp gradient between 
the cold cloud and hot ICM over dynamically long times \citep[e.g.,][]{Vikhl2002}. 

The field-line stretching should naturally produce pressure anisotropy. 
In \Secref{sec:anis}, we made simple estimates of the degree of pressure 
anisotropy for a sonic flow of plasma. These depend on three parameters: 
the collisional mean free path $\lambda$ and the characteristic scales of 
the flow $L_u$ and parallel temperature gradient $L_T$. Let us now estimate the typical electron anisotropy induced 
by the magnetic-field evolution at the interface of a cold front. Because 
the subcluster is moving at around the virial speed, the flow of the hot 
ICM in the comoving frame is nearly sonic. From equation~(\ref{eq:DB}) with 
$u\sim v_{{\rm th},i}$, we get $\Delta_B \sim (1/40) \lambda / L_u\sim2 
\times 10^{-3}$, where for cold fronts we took the electron mean free path 
$\lambda \sim 20$ kpc ($T\sim 8$ keV) and the flow scale $L_u\sim 200$ kpc 
[of order the size of the subcluster, e.g., A3667 \citep{MarkevitchVikhl2007}]. 
The degree of polarization is a few times smaller than the anisotropy 
level, because the coherently anisotropic plasma occupies only a fraction of 
volume of the X-ray emitting ICM, and the magnetic-field direction is not 
necessarily perpendicular to the line of sight. Below, we investigate the amount of electron 
anisotropy, the corresponding bremsstrahlung polarization, and their spatial 
patterns in cold fronts first by means of the simplest analytical model of 
magnetic-field-line draping (Section~\ref{sec:analytic}), and then by numerical 
MHD simulations of cold fronts with anisotropic thermal conduction 
(Section~\ref{sec:num}). We also estimate the total anisotropy 
for the simulated cold fronts in Appendix~\ref{app:A}.

\subsection{Analytical model of magnetic-field-line draping}
\label{sec:analytic}

The problem of the stationary MHD flow of a plasma with a frozen-in 
magnetic field around a spherical body was first solved analytically 
by \cite{BernikovSemenov}. They disregarded the magnetic-field 
back-reaction and assumed a velocity field described by the potential 
flow of an incompressible irrotational fluid around a sphere. Here we 
briefly summarize their derivation and use the resulting magnetic field 
near the body to calculate electron pressure 
anisotropy and thermal bremsstrahlung polarization.
 
\begin{figure}
\centering
\includegraphics[width=84mm]{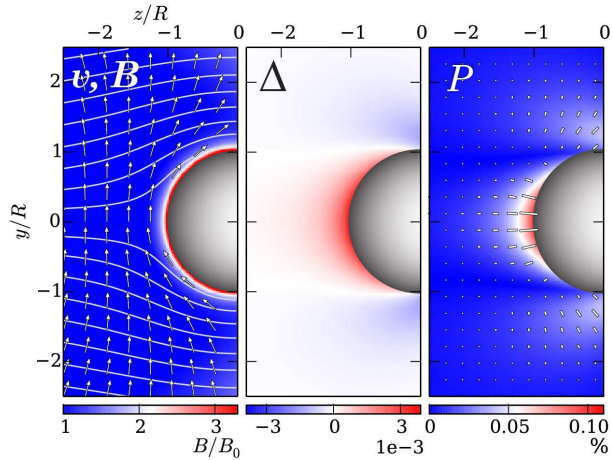}
\caption[Generation of pressure anisotropy and thermal bremsstrahlung 
polarization during kinematic draping of magnetic-field lines around 
a spherical body]
{Generation of pressure anisotropy and thermal bremsstrahlung 
polarization during kinematic draping of magnetic-field lines around 
a spherical body. The velocity field is the potential flow of an 
incompressible fluid around a sphere [equation~(\ref{eq:upot})]. 
The magnetic field is an approximate solution of the ideal 
kinematic MHD equations near the sphere [equations~(\ref{eq:BernSem})].
The left and middle panels are the central $yz$ cross sections. 
{\it Left:} the magnetic-field strength $B$ (color) in the units 
of $B$ at infinity, $B_0$; superimposed are the velocity field 
stream lines (contours) and the unit vectors in the direction 
of the magnetic field (arrows). {\it Middle:} the electron pressure 
anisotropy generated by stretching of the field lines by the 
flow. {\it Right}: the degree (color) and direction (line 
segments) of the polarization of thermal bremsstrahlung as seen 
along the line of sight coincident with the $x$ axis).}
\label{fig:Ban}
\end{figure}

In spherical coordinates with the origin at the center of the sphere 
of radius $R$ and the $z$ axis antiparallel to the fluid velocity 
$v_0$ at infinity, the potential flow around the sphere is 
\beq
\label{eq:upot}
\vc{v} = \vc{e}_r \left ( \frac{R^3}{r^3} - 1 \right ) v_0 \cos{\theta} 
		+ \vc{e}_{\theta} 
		\left ( \frac{R^3}{2r^3} + 1 \right ) v_0 \sin{\theta}. 
\eeq
The magnetic field is obtained by solving the stationary ideal 
MHD equations 
\beq
\nabla \times (\vc{v} \times \vc{B})=0, \ \ \nabla \cdot \vc{B} = 0,
\eeq
with a homogeneous magnetic field $B_0$ along the $y$ axis at 
infinity in the left-half space ($z<0$) as the boundary condition. 
We are primarily interested in the approximate solution near the 
sphere ($r-R \ll R$), where stretching of the field lines is 
greatest. 
It reads
\bea
\nonumber
B_r &=& \frac{2}{3}B_0 \sqrt{3(r/R - 1)} \frac{\sin{\theta}}
						{1+\cos{\theta}} \sin{\phi},\\
\nonumber
B_{\theta} &=& B_0 \frac{\sin{\phi}}{\sqrt{3(r/R - 1)}},\\ 
B_{\phi} &=& B_0 \frac{\cos{\phi}}{\sqrt{3(r/R - 1)}}.
\label{eq:BernSem}
\eea
The velocity and magnetic fields are shown in the left panel of 
Fig.~\ref{fig:Ban}. 

We can now apply equation~(\ref{eq:tot_anis}), where only the 
first term on the right-hand side has to be kept, to calculate the 
electron anisotropy $\Delta$. The term $\Delta_n$ is zero because 
the flow is incompressible. The heat-flux contribution to electron 
anisotropy $\Delta_T$ is also zero because in our configuration, 
the cold cloud is completely isolated from the hot ambient plasma 
by the draped field lines. We choose to ignore temperature 
variations of the incompressible gas outside the sphere and assume 
homogeneous temperature, because these variations would otherwise be clearly 
overstated due to the artificial assumption of constant density. 
Then 
\beq
\Delta = \Delta_B = \frac{\gamma}{\nu_{e}}.
\eeq
From the induction equation, we obtain the rate of stretching of the 
field lines $\gamma$:
\beq
\label{eq:stretch}
\gamma \equiv \frac{1}{B}\frac{dB}{dt} = \vc{b} \vc{b} : \nabla \vc{v},
\eeq
where $\vc{b}=\vc{B}/B$ is the unit vector in the direction of the 
magnetic field. The electron collision frequency \citep{Spitzer1962} 
in a hydrogen plasma is
\beq
\label{eq:nu_e}
\nu_e \approx 3 \times 10^{-6}~{\rm yr}^{-1} \left ( 
		\frac{T_e}{8~{\rm keV}} \right )^{3/2} 
		\left ( \frac{n_e}{10^{-3}~{\rm cm}^{-3}} \right )^{-1}. 
\eeq
We take the radius of the sphere $R=200$ kpc, temperature $T_0=8~
{\rm keV}$ and particle density $n_0=10^{-3}~{\rm cm}^{-3}$ as 
fiducial parameters. Let us set the flow velocity at infinity $v_0$ 
to the speed of sound $c_{s0}= (\gamma_{\rm gas} p_0/\rho_0)^{1/2} 
\approx 1400$ km/s. Combining equations~(\ref{eq:upot})--(\ref{eq:nu_e}), 
we can calculate the electron anisotropy $\Delta$. It is shown in the 
middle panel of Fig.~\ref{fig:Ban} in the central $yz$ cross section. 
Its value agrees well with the previous qualitative estimate in 
Section~\ref{sec:cf_qual}.

The final step is to obtain the polarization of thermal bremmstrahlung 
from the linear approximation (\ref{eq:linpoldeg}), which is, indeed, satisfied for our typical values of 
electron anisotropy $\Delta \sim (1-5) \times 10^{-3}$. We only 
consider the case of sufficiently energetic photons ($\e \sim 2-3~T_e$) 
in the hard X-ray range to ignore the photon-energy/electron-temperature 
dependence and assume $G\sim 1$ in equation~(\ref{eq:linpoldeg}). To 
obtain the polarization map, we need to integrate equation~
(\ref{eq:linpoldeg}) weighted with the thermal bremsstrahlung 
emissivity $\kappa_{\rm br}$, 
\beq
\label{eq:emiss}
\kappa_{\rm br} \propto n_e^2 T_e^{-1/2} {\rm exp}(-\e / T_e),
\eeq  
along the line 
of sight \footnote{The exact temperature dependence may 
be slightly different if the correct form of the gaunt factor is 
adopted, but it practically does not affect our results 
because of small temperature variations outside of the cold cloud.}, taking into account the rotation of the polarization 
vectors due to the changing orientation of the magnetic field. 
Here and in what follows, we simply assume that the X-ray emitting volume is restricted to a 
cubic region of size $5R=1$ Mpc, ignoring the effects of the geometry of the host cluster and 
the location of the cold front inside the host cluster. This should not 
change our results qualitatively, only introducing a large-scale factor of order unity to the 
anisotropy and polarization degree. 
Because the plasma flow is incompressible in our toy model, and its temperature is taken to be homogeneous, 
the emissivity is a constant outside the sphere. We choose the 
$x$ axis as the direction of the line of sight, because in this 
direction, the polarization is greatest, and $xy$ as the reference 
plane. 
We have to integrate both of the independent linear polarization 
types, $P_1$ and $P_2$, where $P_1$ is the polarization measured 
in the vector basis ($\vc{\hat{e}}_y$, $\vc{\hat{e}}_z$), and $P_2$ measured 
in the basis rotated by $\pi/4$ from ($\vc{\hat{e}}_y$, $\vc{\hat{e}}_z$). 
The local polarizations $P_{1,{\rm loc}}$ and $P_{2,{\rm loc}}$ 
relative to the reference plane $xy$ are expressed in terms of  
polarization $P_{B,{\rm loc}}$ [equation~(\ref{eq:linpoldeg})] 
relative to the plane spanned by the local magnetic-field direction and the 
line of sight ($x$ axis):
\bea
P_{1,{\rm loc}} &=& P_{B,{\rm loc}} \cos (2 \zeta),\\
P_{2,{\rm loc}} &=& P_{B,{\rm loc}} \sin (2 \zeta),
\eea 
where $\zeta$ the angle between the projection of the magnetic field 
onto the $yz$ plane and the $y$ axis. Using equation~(\ref{eq:linpoldeg}) 
for $P_{B,{\rm loc}}$ and integrating the local polarization along the 
line of sight, we get
\bea
\label{eq:simP1}
P_1 &=& \frac{\int \Delta \sin^2 \theta \cos (2 \zeta) \ \kappa_{\rm br} dx}
		 {\int \kappa_{\rm br} dx},\\
\label{eq:simP2}
P_2 &=& \frac{\int \Delta \sin^2 \theta \sin (2 \zeta) \ \kappa_{\rm br} dx}
		 {\int \kappa_{\rm br} dx},
\eea 
where $\theta$ is the angle between the local magnetic field 
and the line of sight [as in equation~(\ref{eq:linpoldeg})]. 
The angles $\theta$ and $\zeta$ can be expressed 
in terms of the components of the unit vector in the direction of 
the field $\vc{b}$:
\bea
\cos 2 \zeta &=& (b_y^2-b_z^2)/(b_y^2+b_z^2),\\
\sin^2 \theta &=& 1-b_x^2.
\eea
The total linear polarization $P$ is 
\beq
P = (P_1^2 + P_2^2)^{1/2}.
\eeq
The polarization position angle relative to $\vc{\hat{e}}_y$ is set by angle 
$\psi$, 
\beq
\label{eq:psi}
\psi = \frac{1}{2} {\rm atan} \frac{P_2}{P_1}.
\eeq
The resulting thermal bremsstrahlung polarization pattern is 
shown in the right panel of Fig.~\ref{fig:Ban}, where color 
indicates the degree of polarization, and line segments the 
position angles [calculated by equation~(\ref{eq:psi})] in the $yz$ plane. 
The characteristic degree of polarization is $\sim 0.1 \%$.
If we integrate the polarization along a line of sight at angle 
$\theta' \ne 0$ to the $x$ axis instead, the effect becomes a 
factor of $\cos^2 \theta'$ smaller from the form of equations~(\ref{eq:simP1}) 
and (\ref{eq:simP2}).

\subsection{MHD simulations of cold fronts}
\label{sec:num}  
\subsubsection{Description of the code and setup}

To simulate cold fronts for the purpose of this work, we use an 
MHD code based on the van Leer integrator combined with the constrained 
transport (CT) approach (see \citealt{Stone2009} for a description of 
the numerical method). Anisotropic thermal conduction was implemented 
via a semi-implicit directionally-split scheme with a monotonized central 
(MC) limiter applied to the conductive fluxes to avoid negative 
temperatures \citep{Sharma2011}. The set of equations solved is
\bea
\lefteqn{
\frac{\partial \rho}{\partial t} + \nabla\cdot (\rho \vc{v}),
}&&\\
\lefteqn{
\frac{\partial \rho \vc{v}}{\partial t} + \nabla\cdot 
\left (\rho \vc{v}\vc{v}- \frac{\vc{B}\vc{B}}{4\pi} \right )
+ \nabla p = \rho \vc{g},
}&&\\
\lefteqn{
\frac{\dd E}{\dd t} + \nabla\cdot \left [ \vc{v} (E+p) - 
\frac{\vc{B}(\vc{v}\cdot\vc{B})}{4\pi} \right ] = \rho \vc{g}
\cdot \vc{v} - \nabla\cdot Q,
}&&\\
\lefteqn{
\frac{\dd \vc{B}}{\dd t} = \nabla\cdot (\vc{v}\vc{B}-
\vc{B}\vc{v}),
}
\eea
where
\bea
p&=&p_{\rm th} + \frac{B^2}{8\pi},\\
E &=& \frac{\rho v^2}{2} + \varepsilon + \frac{B^2}{8 \pi},\\
Q &=& -\kappa_{\para} \vc{b}\vc{b}:\nabla T,
\eea
where $p_{\rm th}$ is the gas pressure, $\varepsilon$ the internal 
energy of the plasma per unit volume, $\vc{g}$ the gravitational acceleration, and 
$Q$ the heat flux along the field lines with parallel thermal 
conductivity $\kappa_{\para}$. The plasma is described by an ideal 
equation of state with $\gamma_{\rm gas}=5/3$ and mean molecular 
weight $\mu=0.6$. We take the fiducial value of $\kappa_{\para}=
\kappa_{\rm Sp}$, where $\kappa_{\rm Sp}$ is the Spitzer thermal 
conductivity for an unmagnetized plasma \citep{Spitzer1962}. We ignore any potential 
mechanisms whereby parallel thermal conduction might be suppressed 
(e.g., magnetic mirrors, \citealt{CC1998,Komarov2016}, or 
electron kinetic instabilities, \citealt{Riquelme2016}), as 
we are looking for an upper estimate of the polarization effect.

We initialize a 3D region of hot dilute plasma ($T_{\rm out}=8$ 
keV, $n_{\rm out} = 10^{-3}$ cm$^{-3}$) of spatial extent $L=1$ 
Mpc with a cold spherical subcluster ($T_{\rm in}=4$ keV) of radius $R=200$ 
kpc embedded at the center. The distribution of density inside the 
cold cloud is described by a beta model,
\beq
n_{\rm in} = n_c[1+(r/r_c)^2]^{-3\beta'/2},
\eeq
with $\beta'=2/3$, core radius $r_c=R/\sqrt{3}\approx 115$ kpc, 
and central density $n_c=8n_{\rm out}$. The gravitational 
acceleration $\vc{g}$ models the effect of a static dark matter 
halo at the center of the computational domain, and is set to 
balance the initial pressure gradient inside the subcluster. 
The edge of the subcluster at $r=R$ is a contact discontinuity: 
the temperature experiences a factor-of-two jump in the direction of 
the hot ambient plasma, while the density decreases by a factor of two to 
keep the pressure continuous. The problem is solved in the frame 
comoving with the subcluster. Initially, the cold cloud is at 
rest, while the velocity of the surrounding gas $v_0$ is set 
to the sound speed in the hot ambient plasma, $c_{s0}=(\gamma_{\rm gas} 
p_{\rm out}/\rho_{\rm out})^{1/2}=(\gamma_{\rm gas} k T_{\rm out} 
/ \mu m_p)^{1/2} \approx 1400$ km/s. This setup is similar to the 
one used by \cite{Asai2007}.

To make estimates of the bremsstrahlung polarization generated 
by electron pressure anisotropy due to stretching of the 
magnetic-field lines and to heat fluxes, we analyze the results 
of two runs with different structure of the magnetic field. 
In both runs, the initial plasma $\beta=200$. The first 
run is initialized with a homogeneous magnetic field along 
the $y$ axis, perpendicular to the cold cloud velocity. The 
initial magnetic field in the second run is random with a 
Gaussian distribution and correlation length $l_B = L/10\approx 
100$ kpc. We note that the statistics of magnetic fields in 
galaxy clusters are unlikely to be Gaussian, and the reported 
values of the correlation length inferred from the Faraday 
rotation observations are about an order of magnitude smaller 
\citep[e.g.,][]{Vogt2005}. Therefore, this run is merely 
illustrative and demonstrates only qualitative differences 
between cases with uniform and tangled magnetic field.

\subsubsection{Results for the case of a homogeneous magnetic 
field}

\begin{figure*}
\centering
\includegraphics[width=180mm]{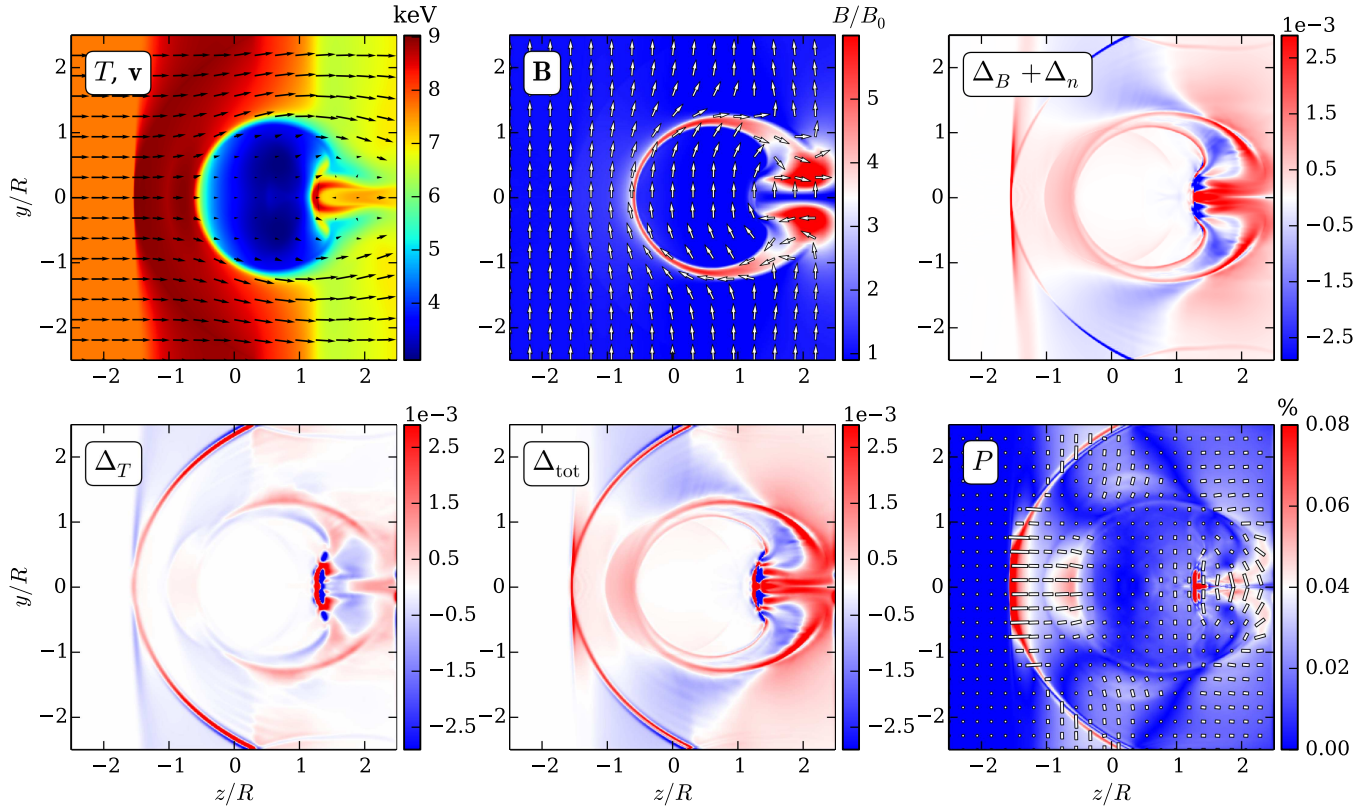}
\caption{A simulation of a cold front with a homogeneous initial 
magnetic field along the $y$ direction. All the panels except 
for the bottom right are the central $yz$ cross sections at time 
$t\approx0.3$ Gyr. The top left panel shows the temperature map 
(color) and the velocity field (arrows). The magnetic field 
$\vc{B}$ is shown in the top middle panel (color: field strength; 
arrows: unit vectors in the magnetic-field direction). The 
different components of the electron anisotropy and the total 
electron anisotropy are demonstrated in the top right, bottom left, 
and bottom middle panels. The bottom right panel shows the 
resulting polarization map in the $x$ direction. }
\label{fig:pol_Bc}
\end{figure*} 

The central $yz$ cross sections of the plasma temperature $T$, 
velocity field $\vc{v}$ and magnetic field $\vc{B}$ are shown 
in the top left and top middle panels of Fig.~\ref{fig:pol_Bc} 
at time $t\approx0.3$ Gyr. The anisotropy pattern at the cold 
front interface is similar to the one in the analytical model 
of the field-line draping (top right panel of Fig.~\ref{fig:pol_Bc}), 
with the typical magnitude of the anisotropy $\Delta\sim 10^{-3}$. Using the 
continuity and induction equations, we can express the degree 
of anisotropy produced by the field-line stretching and 
compression of the gas as
\beq
\label{eq:B_n_Bc}
\nonumber
\Delta_B+\Delta_n = \frac{1}{\nu_e} \lt (\vc{b}\vc{b}:\nabla \vc{v} -\frac{1}{3} \nabla\cdot\vc{v} \rt ).
\eeq
Compression contributes via the divergence of the velocity field 
$\nabla\cdot\vc{v}$. Because $\nabla\cdot\vc{v}$ is positive ahead 
of the subcluster, the electron anisotropy at the interface is 
reduced compared to the incompressible model. The sharp boundary 
of the anisotropy pattern ahead of the front is due to a 
discontinuity in the static gravitational acceleration, which is set to zero 
outside the sphere of radius $R$. This does not affect our estimate 
of the degree of polarization and of the size of the polarized region. 
Because the magnetic field points in the $y$ direction initially, 
the heat flux across the interface is fully suppressed, while in the
regions where the orientation of the field lines is not perfectly 
perpendicular to the temperature gradients, the heat flux 
contribution is noticeable (see the bottom left panel of Fig.~
\ref{fig:pol_Bc}).      

The new features of the simulated cold front, compared to 
the simplistic analytical model studied in Section~\ref{sec:analytic}, 
are the presence 
of a weak bow shock in front of the subcluster and the formation 
of turbulent vortices that efficiently amplify the magnetic field 
behind the subcluster. Let us analyze them in more detail. 

At the moment of taking the snapshot shown in Fig.~\ref{fig:pol_Bc} ($t\approx0.3$ Gyr), the 
bow shock is slowly receding from the cold front at speed $u_{\rm sh} 
\approx 250$ km/s. Let us first discuss the contribution to the anisotropy 
at the shock brought in by the compression of the gas. The source 
of the anisotropy is the jump of the 
normal velocity and tangential component of the magnetic field at 
the shock close to the $z$ axis. The passage of the shock amplifies 
the $y$ component of the magnetic field in the downstream flow.
From equation~(\ref{eq:B_n_Bc}) with $\vc{b}\vc{b}:\nabla \vc{v}=0$ 
(close to the $z$ axis, the velocity only changes in the direction perpendicular to the field 
lines), we can estimate the anisotropy $\Delta_B+\Delta_n$ 
at the shock:
\beq
\label{eq:sh_anis}
\Delta_{B,\rm sh}+\Delta_{n,\rm sh} \sim -10^{-2} \frac{v_{z,{\rm d}}-v_{z, {\rm u}}}{c_{s0}} \frac{\lambda}{\delta},
\eeq
where $v_{z,{\rm u}}$ and $v_{z,{\rm d}}$ are the up- and downstream 
normal velocities, $\delta$ the width of the shock, and $\lambda$ 
the electron mean free path. The normal velocity discontinuity 
contributes to the electron anisotropy via the non-zero velocity 
divergence. The upstream velocity is the speed of sound, $v_{z,{\rm u}}=
v_0=c_{s0}$, while the normal velocity jump in the frame of the shock 
from the Rankine-Hugoniot conditions (consider the magnetic field 
dynamically unimportant) is
\beq
\frac{v_{z,{\rm d}}+u_{\rm sh}}{v_{z,{\rm u}}+u_{\rm sh}} = \frac{(\gamma_{\rm gas}+1)M_1^2}
								{(\gamma_{\rm gas}-1)M_1^2+2}\approx 0.8,
\eeq
where $M_1=(v_{z,{\rm u}}+u_{\rm sh})/c_{s0} \approx 1.18$ is 
the Mach number of the upstream gas in the frame of the shock. 
Then we can infer the velocity jump in laboratory frame $v_{z,
{\rm d}}/v_{z, {\rm u}} \approx 0.8$. Taking the shock width 
$\delta \sim \lambda$, from equation~(\ref{eq:sh_anis}) we 
estimate the typical value of anisotropy at the shock $\Delta_
{B+n,\rm sh} \eqsim 2\times10^{-3}$. Results of the numerical 
simulations agree well with this estimate (see the top right 
panel of Fig.~\ref{fig:pol_Bc}). At angles larger than $\pi/4$ 
from the $z$-axis, the term $\vc{b}\vc{b}:\nabla \vc{v} \approx 
b_y^2 \partial_{y} v_y<0$ starts to dominate at the shock, because 
there is a jump in the $y$-velocity parallel to the field lines, 
and the magnetic field is compressed along the $y$ direction 
producing negative electron anisotropy.  

Close to the $z$ axis, the magnetic field is perpendicular to the 
temperature gradient, and there is no heat flux across the shock. 
However, away from the $z$ axis, the magnetic field only partly 
impedes thermal conduction. Although the strong parallel conductivity 
smears the temperature gradient, a small jump of the temperature and its gradient 
along the shock is still left behind. The jump $\delta T 
/ T$ is of order $0.5~\%$, and the level of positive anisotropy 
it generates is of the same order, because, from equation~(\ref{eq:DT}), 
$\Delta_T\sim (\lambda/L_T)^2 \delta T/T$ (we took $L_u=L_T$ because 
the heat flux changes at the scale of the shock width, as well as 
the temperature). The scale of the gradient $L_T$ is of the order of the 
shock width, which can be approximated by the mean free path 
$\lambda$. Then the anisotropy is simply $\Delta_T \sim \delta 
T/T \sim 0.5~\%$. This is seen in the bottom left panel of Fig.~
\ref{fig:pol_Bc}. 

Another notable feature of the simulated cold front is the 
amplification of the magnetic field behind the subcluster (top 
middle panel of Fig.~\ref{fig:pol_Bc}), previously reported by 
\cite{Asai2007}. The amplification is caused by stretching 
of the field lines along the $z$ direction by the vortices 
generated by the flow of the ambient gas around the subcluster. 
The magnetic field is amplified more efficiently than at the cold 
front interface, because the vortices are smaller than the subcluster, 
and thus produce a velocity strain rate larger than that ahead of 
the subcluster by a factor of the ratio of the subcluster size to 
the size of the vortices. Therefore, they are expected to generate 
more electron anisotropy, which is clearly seen in the top right 
panel of Fig.~\ref{fig:pol_Bc}. 

The total electron anisotropy is shown in the bottom middle panel of 
Fig.~\ref{fig:pol_Bc}. The corresponding polarization of thermal 
bremsstrahlung is calculated by equations~(\ref{eq:emiss})--(\ref{eq:psi}), 
now taking account of the spatial variation of the bremsstrahlung 
emissivity [equation~(\ref{eq:emiss})], and demonstrated in the bottom 
right panel of Fig.~\ref{fig:pol_Bc}. The polarization is generally 
dominated by stretching of the field lines and the compressibility 
effects. There are three regions where the degree of polarization is 
at $\sim 0.1\%$ level: 1) at the cold-front interface due to stretching 
of the field lines in the $y$ direction; 2) at the bow shock close to 
the $z$ axis due to the compressibility term in equation~(\ref{eq:B_n_Bc}); 
3) behind the subcluster due to amplification of the magnetic field 
along the $z$ direction by the turbulent vortices. 

\subsubsection{Results for the case of a random magnetic field}

\begin{figure*}
\centering
\includegraphics[width=180mm]{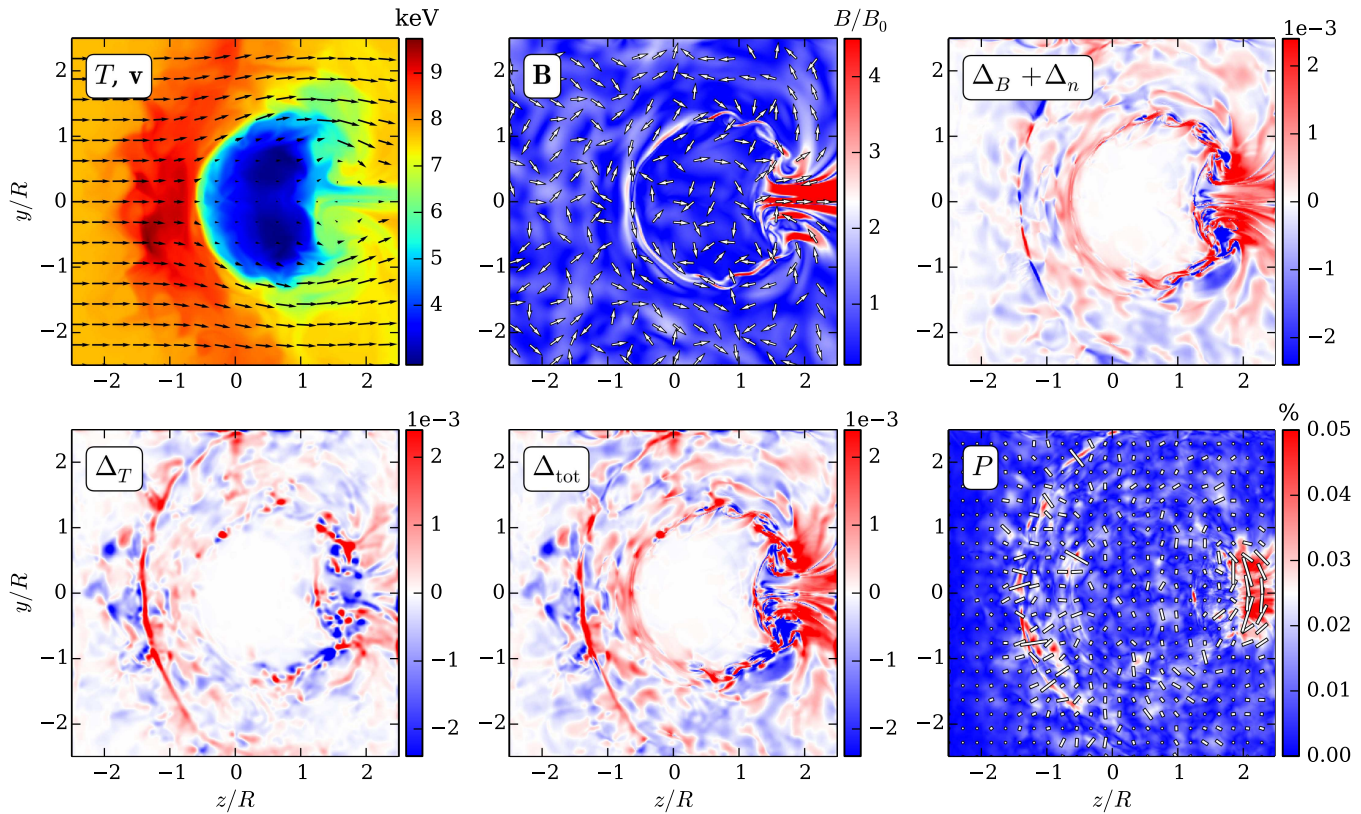}
\caption{A simulation of a cold front with a random Gaussian initial 
magnetic field with correlation length $l_B\approx100$ kpc. The 
panels show the same quantities as in Fig.~\ref{fig:pol_Bc}.}
\label{fig:pol_Br}
\end{figure*}

It is currently believed (based on numerical and indirect observational evidence) that the ICM is turbulent (see, 
e.g., \citealt{Inogamov2003, Schuecker2004, Schek2006turb, 
Subramanian2006, Zhur2014}), and, therefore, the magnetic fields in the ICM are 
tangled by random motions of the plasma. Here we model the effect of  
random topology of the field lines by generating a random Gaussian 
magnetic field with correlation length $l_B = 100$ kpc. The mean plasma 
$\beta=2 p_{\rm out}/\langle B^2 \rangle=200$, where $\langle B^2 \rangle 
= B_0^2$. Analogously to the case of a homogeneous magnetic field, 
the results of our simulations are shown in Fig.~\ref{fig:pol_Br}. The 
random field topology diminishes the electron anisotropy 
produced by stretching of the field lines at the cold-front interface 
(top right panel of Fig.~\ref{fig:pol_Br}) and almost completely 
wipes out its contribution to the total polarization (see the bottom 
right panel). We should remark that due 
to numerical diffusivity, during compression and stretching of the 
field-line loops by the convergence flow at the front, the opposite 
orientations of the field may reconnect, thus modifying the field-line 
topology in the region where one expects to see electron anisotropy. 
Therefore, our numerical estimate in this region might be understated. 

Because now the field-line orientation at the bow shock is random, 
in addition to the compression term [equation~(\ref{eq:sh_anis})], 
the heat fluxes also positively contribute to the total anisotropy 
everywhere across the shock (see the bottom left panel of Fig.~
\ref{fig:pol_Br}). The level of electron anisotropy generated by the 
turbulent vortices behind the shock is practically unchanged compared 
to the simulation with a homogeneous magnetic field. In a random 
magnetic field, this mechanism appears to be the most efficient. 

The resulting polarization map (bottom right panel) indicates 
that 1) polarization at the cold-front interface is practically 
indiscernible; 2) the degree of polarization at the shock is 
$\sim 0.05\%$; 3) the largest polarization, $\sim 0.1\%$, is 
achieved behind the cold cloud via the magnetic-field amplification 
by the turbulent backflow. 

\section{Discussion}
\label{sec:disc}

At present, the only astrophysical object from which a polarized 
signal has been reliably detected in X-rays below 10 keV is the 
Crab Nebula \citep{Weisskopf1978}, dating back to the 70s. The 
progress with the development of the new generation of X-ray 
polarimeters \citep[e.g.,][]{Soffitta2013,Weisskopf2013,Jahoda2014} has led to a dramatic 
increase of the expected sensitivity that could open a new 
observational window into a variety of astrophysical objects. 
Magnetars, radio and accreting X-ray pulsars, reflected radiation 
in X-ray binaries or AGNs are all among the promising targets 
for missions like XIPE, IXPE and PRAXyS.

In the majority of those objects, polarization is either associated 
with non-thermal emission (e.g., synchrotron radiation of relativistic 
electrons) or with scattering in aspherical geometries. Here we discuss 
the polarization of {\it thermal} emission from the hot gas in galaxy 
clusters. This is an interesting question for at least two reasons: (i) 
clusters of galaxies are considered as possible unpolarized targets 
for calibration purpose and (ii) weak polarization of thermal 
bremsstrahlung potentially could serve as a proxy for the plasma 
properties on extremely small scales, not directly resolvable with 
the current or future X-ray missions.

As we showed in the previous sections, the polarization of thermal 
bremsstrahlung naturally arises from the anisotropy of the electron 
distribution function driven by stretching of the magnetic-field lines 
and/or temperature gradients along the field. Pitch-angle scattering of 
the electrons controls the level of anisotropy, and even if it is set 
purely by Coulomb collisions, the anisotropy is always small. Further reduction of 
the observed polarization signal is expected if many uncorrelated 
regions with varying orientation of anisotropy are present along the 
line of sight, leading to effective averaging of the signal. From this 
point of view, the most promising are the configurations with a 
large-scale flow that provides a coherence of 
structures and drives the anisotropy. Our 
qualitative estimates show that the expected degree of polarization is close to 
0.1\% for rather idealized configurations that exhibit shocks and cold 
fronts. 

It is worth noting that aside from polarizing thermal bremsstrahlung, electron 
pressure anisotropy is also capable of producing a small
degree of polarization of the Sunyaev-Zeldovich (SZ) signal \citep{SZ1980} and 
the emission in collisionally-excited X-ray lines (e.g., He-like triplets of 
silicon, sulphur and iron, see \citealt{Palchik1995}, p. 140, and references 
therein). However, these two effects are both subjected to somewhat higher 
possible contamination coming from the Thomson scattering of cluster central 
AGN radio emission and the contribution of the kinetic SZ effect in the first case 
\citep{SZ1980,SS1999,Diego2003}, and resonant-scattering-induced polarization 
in the latter case \citep[][]{Sazonov2002,Zhuravleva2010}.

The small degree of polarization makes galaxy clusters a suitable (unpolarized) calibration target 
for the forthcoming generation of X-ray polarimeters. We note in passing 
that two other mechanisms could also contribute to the polarization of 
thermal emission of the hot gas in clusters. One is Thomson 
scattering of centrally concentrated X-ray emission by the electrons; 
another is resonant scattering of emission-line photons 
\citep[][]{Sazonov2002,Zhuravleva2010}. Both effects have a clear signature of the 
polarization plane being perpendicular to the direction towards the cluster 
center and are expected to disappear if the integrated signal (over a 
circular region around the cluster center) of a relaxed cluster is used. 
Even if an offset region is considered, one can crudely estimate the 
expected level of polarization. Given that the Thomson optical depth in 
clusters is at the level of $10^{-3}$, the scattered thermal emission 
should not be polarized by more than a fraction of this value. For resonant lines, 
the optical depth is larger, but the effect is confined to line photons and does 
not affect the continuum. On the whole, clusters are suitable calibration 
objects for IXPE, XIPE, or PRAXyS.

Nevertheless, if future polarimeters with capabilities well 
beyond currently developed instruments could detect polarization from 
carefully selected clusters with large-scale substructure, it would 
imply that one has a way to constrain the effective collisionality of 
electrons. Of course, clusters of galaxies are not the only objects where 
polarization of thermal emission could be present. As an example one 
could consider hot radiatively inefficient flows around black holes or 
neutron stars that might have conditions suitable for generation of 
sufficient electron anisotropy. We defer this question to 
further studies.

\section{Conclusions}

We have studied the effect of polarization of thermal bremsstrahlung 
emission in a weakly collisional astrophysical plasma due to (small) 
electron pressure anisotropy. Stretching of magnetic-field lines 
by a flow of plasma, compression/rarefaction, or heat fluxes all lead 
to generation of pressure anisotropy in a plasma where the Larmor radii 
of the charged particles are much smaller than their mean free paths. In 
the case of ordered plasma motions with a certain preferred direction, 
electron anisotropies may produce polarization of thermal bremsstrahlung 
emission. The degree of polarization is a few times lower than the 
anisotropy level (depending on the size of the region of coherently 
anisotropic electrons and on the bremsstrahlung photon energy). 

We have estimated the upper bounds on the degree of polarization in cold fronts 
in the ICM as they represent a perfect example of converging flows 
and large temperature gradients. Cold fronts may also be associated 
with additional features (although not always observed), such as bow 
shocks or turbulent vortices generated behind subclusters. We have 
found that a small polarization, at $\sim0.1\%$ 
level, can be generated by either the converging flow, weak bow 
shock, or vortices behind the cold front. Although at the moment, 
such a small degree of bremsstrahlung polarization at energies of a 
few $kT$ cannot be observed, future observations of this effect might 
provide a valuable insight into the generation of pressure anisotropies in 
astrophysical plasmas. The absence of polarization at the estimated 
level could also set lower limits on electron collisionality in 
the ICM, which may be enhanced by scattering off microscale magnetic 
fluctuations \citep{Riquelme2016} or magnetic trapping by the mirror instability at the 
scale of the ion Larmor radius \citep{Komarov2016}.     

\label{sec:concl}

\section*{Acknowledgements}

The authors thank M. W. Kunz and A. Spitkovsky for useful discussions. 
EC acknowledges partial support by grant No. 14-22-00271 from the Russian Scientific Foundation. 
AAS's work was supported in part by grants from the UK STFC and EPSRC.


\bibliographystyle{mn2e}
\bibliography{bibliography}

\appendix 

\section{Total anisotropy and microscale instabilities}
\label{app:A}

\begin{figure*}
\centering
\includegraphics[width=175mm]{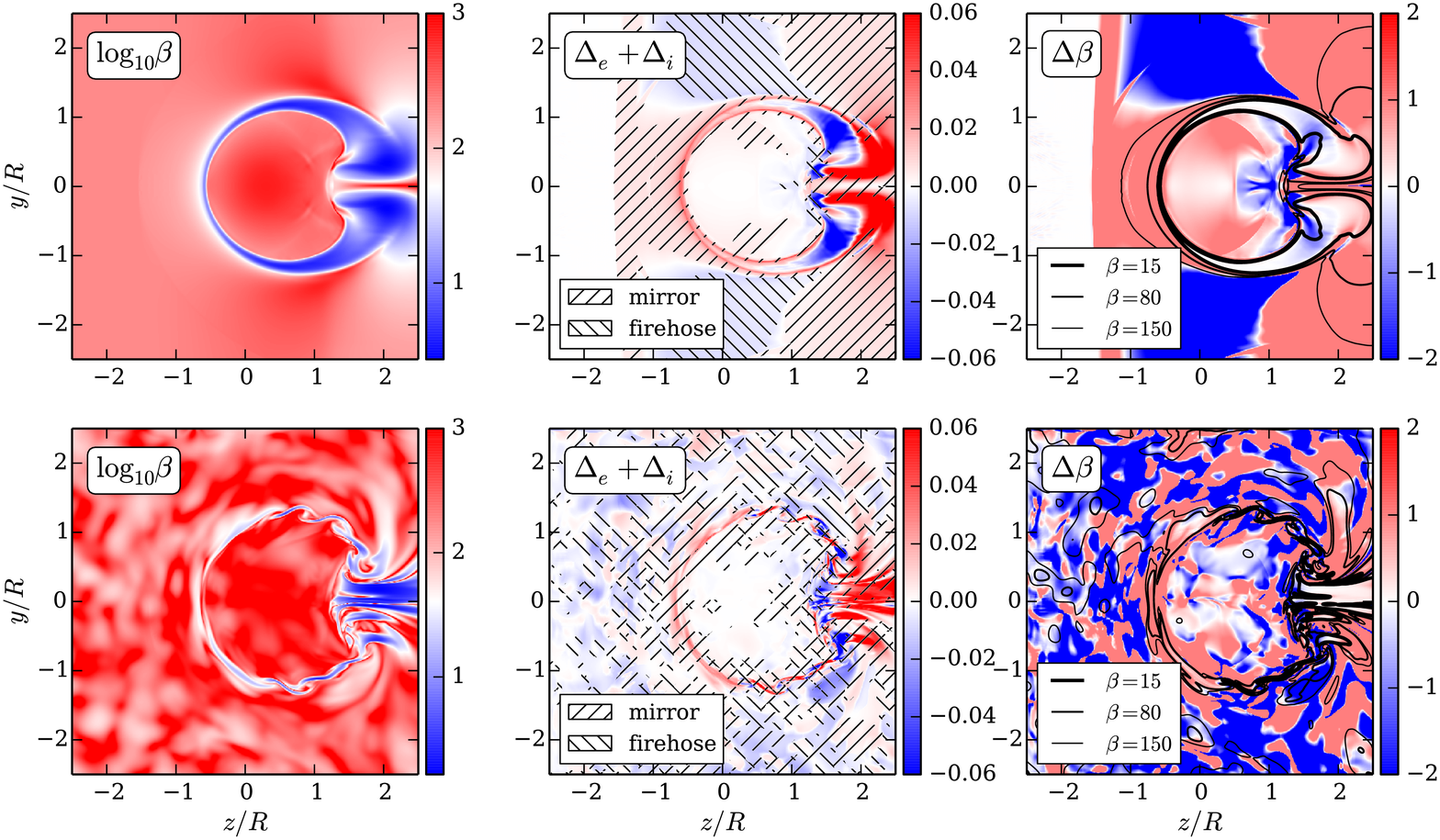}
\caption{The total (ion+electron) anisotropy for the simulated cold fronts. {\it Left panels}: 
the plasma $\beta$. {\it Middle panels}: the total anisotropy $\Delta_e+\Delta_i$, unstable 
regions are hatched; in the unstable regions, the plasma is kept marginal: 
$\Delta_e+\Delta_i \approx 1/\beta$ for the mirror, $\Delta_e+\Delta_i = -2/\beta$ for the 
firehose instabilities. {\it Right panels}: the total anisotropy multiplied by $\beta$ with 
contours of $\beta$ overlaid; this quantity is equal to 1 in the mirror-unstable regions, and -2 in the 
firehose-unstable regions.}
\label{fig:tot}
\end{figure*}

Here, for illustrative purposes, we calculate the total anisotropy $\Delta_e+\Delta_i$ by 
equation~(\ref{eq:tot_anis}) for the two 
simulated cold fronts with homogeneous and random magnetic fields. As it is seen in 
Fig.~\ref{fig:tot}, the ion contribution to the anisotropy is large enough to trigger the firehose and 
mirror instabilities all over the computational domain. 
The plasma is rendered unstable when $|\Delta_e+\Delta_i| \gtrsim 1/\beta$. The instabilities 
maintain the plasma in the marginal state: $\Delta_e+\Delta_i \approx 1/\beta$ for the mirror 
(the general mirror instability condition reduces to this simple expression when both the ion and 
electron anisotropies are small: $\Delta_e \ll 1$, $\Delta_i\ll 1$), 
$\Delta_e+\Delta_i=-2/\beta$ for the firehose instabilities.
In some regions, $\beta$ becomes low and allows a rather high level 
of anisotropy, while keeping the plasma stable. In the case of a homogeneous magnetic field, this is clearly seen 
in the low-$\beta$ layer around the cold front and behind the front, where the 
magnetic field is amplified by turbulent eddies. In the case of a random magnetic field, 
stable regions of high anisotropy form behind the front.

\section{Relativistic bremsstrahlung differential cross sections}
\label{app:B}

The expressions for the relativistic bremsstrahlung differential 
cross sections were first given by \cite{Gluckstern}. After being 
rearranged in a more convenient way \citep{BaiRamaty} and corrected 
for typos, they read
\smallskip
\beq
\label{eq:dsigmaperp}
\dsigmaperp=A\left(B_{\perp}+\frac{L}{p\pd}C_{\perp}+\frac{l_0}{\pd Q}D_{\perp}-\frac{E}{p^2\sin^2\theta}\right),
\eeq
\beq
\label{eq:dsigmapara}
\dsigmapara=A\left(\tilde{ B}_{\para}(\theta)+\frac{L}{p\pd}\tilde{ C}_{\para}(\theta)+\frac{l_0}{\pd Q}D_{\para}+\frac{E}{p^2\sin^2\theta}\right),
\eeq
\smallskip
\beq
\label{eq:A}
A=\frac{Z^2}{8\pi}\frac{{r_0}^2}{137}\frac{\pd}{p}\frac{1}{\e},
\eeq
\beq
\label{eq:Bperp}
B_{\perp}=-\frac{5\g^2+2\ggd+1}{p^2\D^2}-\frac{p^2-k^2}{Q^2\D^2}-\frac{2k}{p^2\D},
\eeq
\smallskip
\beq
\label{eq:Bpara}
B_{\para}=-\frac{5\g^2+2\ggd+5}{p^2\D^2}-\frac{p^2-k^2}{Q^2\D^2}+\frac{2(\g+\gd)}{p^2\D}-\frac{4l}{\pd\D},
\eeq
\beq
\label{eq:BparaT}
\tilde{B}_{\para}(\theta)=B_{\para}+\frac{8(2\g^2+1)}{p^2\D^4}\sin^2\theta,
\eeq
\smallskip
\bea
\nonumber
C_{\perp}&=&\frac{2\g^2(\g^2+\gd^2)-(5\g^2-2\ggd+\gd^2)}{p^2\D^2}\\
\label{eq:Cperp}
		&&+\frac{k(\g^2+\ggd-2)}{p^2\D},
\eea
\bea
\nonumber
C_{\para}&=&\frac{2\g^2(\g^2+\gd^2)-(9\g^2-4\ggd+\gd^2)+2}{p^2\D^2}\\
\label{eq:Cpara}
	&&+\frac{k(\g^2+\ggd)}{p^2\D},
\eea
\beq
\label{eq:CparaT}
\tilde{C}_{\para}(\theta)=C_{\para}+\frac{4\g(3k-p^2\gd)}{p^2\D^4}\sin^2\theta, 
\eeq
\smallskip
\beq
\label{eq:Dperp}
D_{\perp}=\frac{k}{\D}-\frac{k(p^2-k^2)}{Q^2\D}+4,
\eeq
\beq
\label{eq:Dpara}
D_{\para}=\frac{4}{\D^2}-\frac{7k}{\D}-\frac{k(p^2-k^2)}{Q^2\D}-4,
\eeq
\smallskip
\bea
\nonumber
E&=&\frac{2L}{p\pd}\left(2\g^2-\ggd-1-\frac{k}{\D}\right)-\frac{4 l_0}{\pd Q}
\label{eq:E}
\left(\D-\gd\right)^2\\
	&&-\frac{2l(\D-\gd)}{\pd},
\eea
\smallskip
where
\beq
\label{eq:defsg}
\g=E/\mec+1 ;~ \gd=\g-\e/\mec ; 
\eeq
\beq
\label{eq:defsp}
p=\sqrt{\g^2-1};~ \pd=\sqrt{\gd^2-1};~ k=\e/\mec;
\eeq
\beq
\label{eq:defsqd}
Q^2=p^2+k^2-2pk\cos\theta;~ \D=\g-p\cos\theta;
\eeq 
\bea
\nonumber
L&=&2\ln\left(\frac{\ggd+p\pd-1}{\ggd-p\pd-1}\right);~ l=\ln\left(\frac{\gd+\pd}{\gd-\pd}\right);\\
\label{eq:defsls}
l_0&=&\ln\left(\frac{Q+\pd}{Q-\pd}\right);
\eea
$r_0=e^2/(m_e c^2) \approx 2.82\times 10^{-13}~{\rm cm}$ is the classical electron radius, 
the rest of the notations were introduced in Section~\ref{sec:pol}.

\end{document}